%% file: main.tex
\documentclass[twocolumn]{article}
\usepackage[utf8]{inputenc}
\usepackage{graphicx}
\usepackage[left=1.5cm,right=2cm,top=2cm,bottom=2cm]{geometry}
\usepackage{authblk}
\usepackage{cite}
\usepackage{float}
\usepackage{subfigure}
\usepackage{multirow}
\usepackage{longtable}
\usepackage{acronym}
\usepackage{hyperref}
\usepackage[T1]{fontenc}
\usepackage{authblk}
\usepackage{amsmath,amssymb}
\usepackage[table,xcdraw]{xcolor}

\newcommand{\msun}{\ensuremath{M_{\odot}}}
\newcommand{\mchirp}{\ensuremath{\mathcal{M}}}
\newcommand{\mcdet}{\ensuremath{\mathcal{M}^\mathrm{det}}}
\newcommand{\deff}{D_\mathrm{eff}}
\newcommand{\deffi}{D_{\mathrm{eff},i}}
\newcommand{\dest}{\tilde{D}_\mathrm{L}}
\newcommand{\dinj}{D_\mathrm{inj}}
\newcommand{\sigmaest}{\tilde{\sigma}_{D_\mathrm{L}}}

\acrodef{CBC}{compact binary coalescence}
\acrodef{BNS}{binary neutron star}
\acrodef{MG}{MassGap}
\acrodef{NSBH}{neutron star-black hole}
\acrodef{BBH}{binary black hole}
\acrodef{NS}{neutron star}
\acrodef{BH}{black hole}
\acrodef{SNR}{signal-to-noise ratio}
\acrodef{O3}{the third observing run}
\acrodef{O3a}{the first half of the third observing run}
\acrodef{O3b}{the second half of the third observing run}
\acrodef{O4}{the fourth observing run}
\acrodef{GraceDb}{Gravitational-wave Candidate Event Database}
\acrodef{GCN}{Gamma-ray Coordinates Network}
\acrodef{FAR}{False Alarm Ratio}
\acrodef{GWTC-2}{the second Gravitational Wave Transient Catalog}
\acrodef{GWTC-3}{the third Gravitational Wave Transient Catalog}
\acrodef{GraceDB}{the Gravitational-Wave Candidate Event Database}
\acrodef{LIGO}{the Laser Interferometer Gravitational-wave Observatory}
\acrodef{KDE}{kernel density estimate}
\acrodef{EM}{electromagnetic}
\acrodef{GW}{gravitational-wave}
\acrodef{PE}{parameter estimation}
\acrodef{LVK}{LIGO-Virgo-KAGRA}

\title{
Astrophysical Source Classification and Distance \\ Estimation for PyCBC Live 
}

\author{Verónica Villa-Ortega}
\author{Thomas Dent}
\author{Andrés Curiel Barroso}
\affil{\textit{Instituto Galego de F\'{i}sica de Altas Enerx\'{i}as, Universidade de Santiago de Compostela, 15782 Santiago de Compostela, Galicia, Spain}}

\date{\today}

\begin{document}

\twocolumn[
  \begin{@twocolumnfalse}
    \maketitle
  \begin{abstract}
    \noindent
    During \acf{O3} of the Advanced LIGO and Advanced Virgo detectors, dozens of candidate \acf{GW} events have been  catalogued.  A challenge of this observing run has been the rapid identification and public dissemination of \acf{CBC} signals, a task carried out by low-latency searches such as PyCBC Live. During the later part of \ac{O3}, we developed a method of classifying CBC sources, via their probabilities of containing neutron star or black hole components, within PyCBC Live, in order to facilitate immediate follow-up observations by electromagnetic and neutrino observatories. This fast classification 
    uses the chirp mass recovered by the search as input, given the difficulty of measuring the mass ratio with high accuracy for lower-mass binaries. We also use a distance estimate derived from the search output to correct for the bias in chirp mass due to the cosmological redshift. 
    We present results for simulated signals, and for confirmed candidate events identified in low latency over \ac{O3}. 
  \end{abstract}
  \vspace{2cm}
  \end{@twocolumnfalse}]
  
\section{Introduction}
%
Since the beginning of observations with the network of advanced \acf{GW} interferometers in 2015, including Advanced LIGO~\cite{LIGOScientific:2014pky} and Advanced Virgo~\cite{VIRGO:2014yos}, low-latency alerts for follow-up by \ac{EM} and other observatories~\cite{VIRGO:2011puf,IceCube:2014yxu,LIGOScientific:2016qac,ANTARES:2017bia,LIGOScientific:2019gag} have been a crucial element in enabling \ac{GW} astronomy as a multi-messenger endeavour.  The detection of joint \ac{GW}-\ac{EM} emission on August 2017 from the merger of two neutron stars~\cite{LIGOScientific:2017vwq} with subsequent emission over the entire accessible \ac{EM} spectrum~\cite{LIGOScientific:2017zic,LIGOScientific:2017ync} dramatically demonstrated the science potential of such follow-up programs.
Despite efforts to increase coverage, the capacities of both terrestrial and space-based observatories to follow up frequent \ac{GW} alerts are limited in speed of pointing, field of view and survey depth.  Therefore, with the expected increasing sensitivity of the \ac{GW} detector network~\cite{KAGRA:2013rdx} and corresponding increase in detection rate, there is an increasing need to distinguish between different classes of signal and source.  Binary systems containing one or two \acp{NS} are of particular interest, in contrast to \ac{BBH}, which are in general not expected to emit electromagnetic or other radiation at merger. 

Binary merger \ac{GW} signals are identified in the data streams from the global detector network almost exclusively by the use of matched filter methods, e.g.~\cite{Allen:2005fk}: see \cite{LIGOScientific:2020ibl,LIGOScientific:2021djp} and references therein for recent results and discussion.\footnote{Exceptions may occur for signals with unusual features which cause a mismatch with commonly used binary coalescence templates, for instance strong amplitude modulation due to orbital plane precession or high orbital eccentricity. 
As yet, there is no clear evidence for such a population of mergers `invisible' to existing templated searches.} We could then hope that the matched filter template(s) that identify a given event would indicate the type of source and the presence of \ac{NS} components.
\begin{figure*}[ht]
    \centering
    \includegraphics[width=\textwidth]{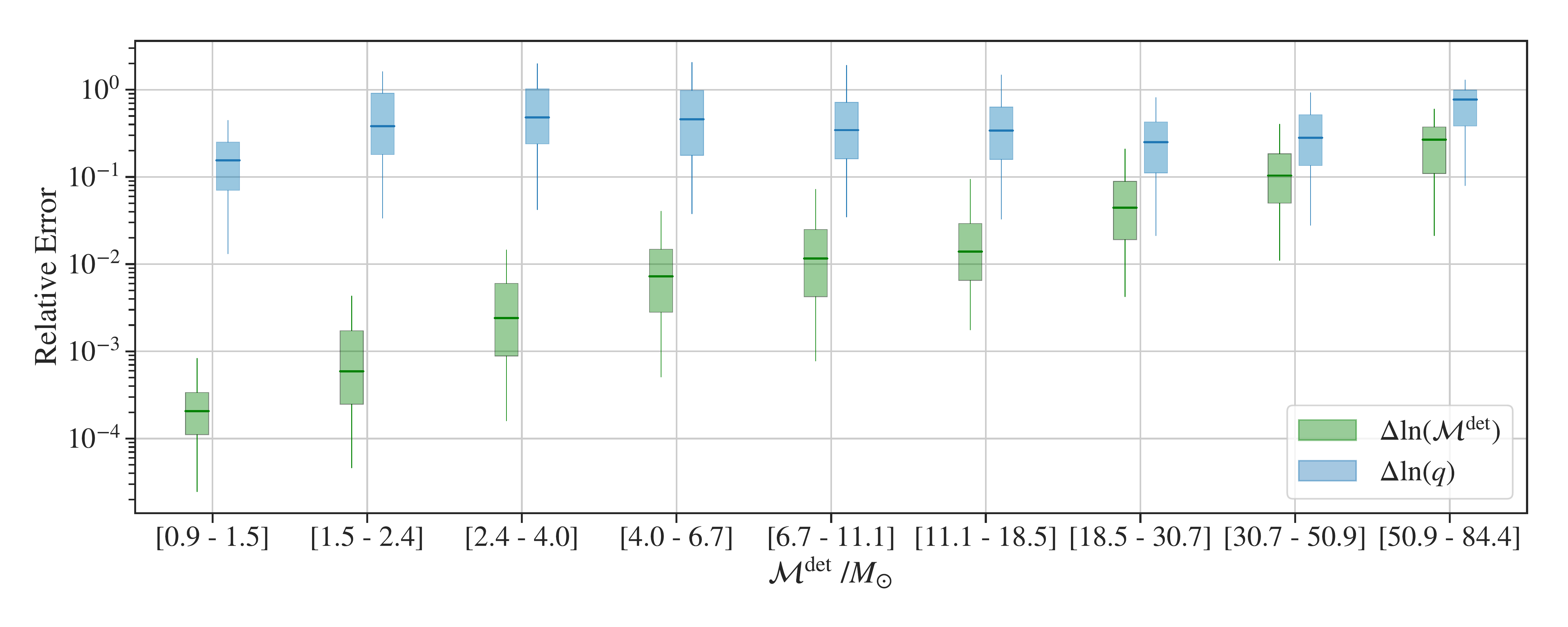}
    \vspace{-0.75cm}
    \caption{Boxplot comparison of redshifted chirp mass $\mcdet$ and mass ratio $q$ relative errors for different detector frame chirp mass intervals. Median values and 50\% and 90\% credible intervals are presented in each case.
    }
    \label{fig:mchirp_rel_err}
\end{figure*}
However, there are several obstacles to deducing the identity of the \ac{GW} source directly from the templated search pipeline output.  While some of these issues are mainly technical, some may be considered fundamental, meaning they may not be overcome by refinements to data analysis methods and can only be addressed (if at all) by increasing detector sensitivity.  The first such fundamental issue is the presence of noise in the detector outputs and the (relatively) low \ac{SNR} of most detected events, of order 10.  This leads us to expect significant uncertainties in any estimate of binary system parameters (e.g.~\cite{Cutler:1994ys}). The second, related issue is the correlation between errors in the system's component spins and in component masses~\cite{Baird:2012cu,Hannam:2013uu} which results in substantial uncertainties in the masses, especially for lower-mass events if a wide range of possible spins is allowed \textit{a priori}.  

A third fundamental issue, which we will not attempt to address, concerns the possibility to distinguish between \ac{NS} and \ac{BH} components \textit{of the same mass} using only \ac{GW} data, via the tidal deformation and disruption of \ac{NS} components before and during merger~\cite{Yang:2017gfb,Brown:2021seh}.  Given current knowledge of the neutron star equation of state and limits on tidal deformability, it seems unlikely that such effects will yield clear information with the current generation of \ac{GW} detectors. 

To distinguish between \ac{NS} and \ac{BH} components, published \ac{LVK} investigations have relied on the estimated masses of the binary components: either imposing an approximate classification boundary, for instance $3\,\msun$ as in
\ac{LVK} public alerts~\cite{EMFollowUserGuide}, or by comparing the estimated masses, including uncertainties due to noise, with the maximum \ac{NS} masses resulting from models of neutron star structure informed by various observations~\cite{LIGOScientific:2021psn}.  Both of these methods use component mass samples resulting from detailed Bayesian parameter estimation analysis (e.g.~\cite{Veitch:2014wba}) which is, so far, computationally prohibitive to pursue in low latency for lower-mass mergers.  For online alerts, more approximate methods have been used~\cite{LIGOScientific:2019gag,EMFollowUserGuide,Chatterjee:2019avs,Kapadia:2019uut} attempting to quantify component mass uncertainties via geometric estimates, via machine learning methods, via the recovered parameters of simulated signals or, in the simplest case, assuming that the template produced by the search gives the true source mass, thus neglecting uncertainties and systematic error.

Here we present a method to estimate both the source classification, i.e.\ the probabilities of either binary component to be \ac{NS} or \ac{BH}, and the source distance directly from the output of the PyCBC Live search pipeline \cite{Nitz:2018rgo,DalCanton:2020vpm}.  The method approximates the uncertainties in source masses by assuming that only one quantity, the chirp mass, is reliably recovered by the search: due to this simplification we are able to provide the estimated distance and classification with $< 50\,$ms additional latency (in comparison to a typical total latency of $\sim 20\,$s including data transfers and time for PyCBC Live to identify candidates~\cite{DalCanton:2020vpm}). 

The paper is structured as follows.  In the next section, we present the technical problem to be addressed and the state of the art in more detail. In Section~\ref{sec:methods} we present the method developed during the later part of O3, and in Section~\ref{sec:injections} we demonstrate the performance of the method on a large set of simulated signals added to real LIGO-Virgo data. In Section~\ref{sec:o3results} we discuss the results of applying the method to \ac{GW} event candidates identified in low latency during O3, and compare these with other source probability estimates. Finally, in Section~\ref{sec:conclusion} we conclude with a summary and discussion of possible future work. 

\section{Context \& Motivation}
\label{sec:motivation}
The search for \acf{CBC} signals in the detector data is performed in low latency by the following matched-filter based analysis pipelines: PyCBC Live, 
GstLAL~\cite{Messick:2016aqy,Sachdev:2019vvd}, MBTAOnline~\cite{Adams:2015ulm} and SPIIR~\cite{76e97cf9801544919973534ed7028b6a}. When a candidate \ac{GW} signal is detected, it is saved in \ac{GraceDB}\footnote{\url{https://gracedb.ligo.org/}} as an \textit{event}. To unify all candidates from the different searches that are neighbours in time, the events collected within a given time window form a \textit{superevent}. One of the candidates is chosen as the \textit{preferred event} based on maximising search \ac{SNR}, for the purposes of issuing public alert information such as sky localization~\cite{EMFollowUserGuide}.

The probability of a \ac{CBC} candidate event of being of astrophysical origin, $p_\mathrm{astro}$, is obtained via a Poisson mixture formalism that models foreground and background event rates and distributions~\cite{Farr:2013yna}; the probability 
of terrestrial noise origin is thus  $p_\mathrm{terr} = 1 - p_\mathrm{astro}$. The latter is distributed in the LIGO/Virgo Public Alerts, along with the probabilities of the candidate event to belong to each of the four \ac{CBC} categories established in \ac{O3}: \ac{BNS}, \acf{MG}, \ac{NSBH} and \ac{BBH}. A detailed description of these categories is given in the next section.
The probabilities $p_\mathrm{BNS}$, $p_\mathrm{MG}$, $p_\mathrm{NSBH}$ and $p_\mathrm{BBH}$, have to sum to $p_\mathrm{astro}$. In the work presented in this paper, we only consider estimation of the \emph{relative} astrophysical probabilities of the four categories.




During the period of \acf{O3}, astrophysical source classification for PyCBC Live candidates was performed by the LIGO/Virgo rapid alert infrastructure using a ``hard cuts'' method, which assigns Boolean weights (either $1$ or $0$) to the different source types based on component mass cuts applied to the reported search template~\cite{GCN25324}. The component masses taken from the point estimates of the search are measured with large uncertainties that are not taken into account in the classification. Another bias and additional source of uncertainty is caused by the fact that template component masses are redshifted with respect to the true masses of the source; this bias is also ignored by the previous method.

A parameter measured by the pipelines with much more accuracy is the combination of component masses known as \textit{chirp mass}:
\begin{equation}\label{eq:mchirp}
\mchirp=\frac{(m_1m_2)^{3/5}}{(m_1+m_2)^{1/5}},
\end{equation}
where $m_1$ refers to the primary component mass in the binary, and $m_2$ refers to the secondary.
Using a set of simulated astrophysical signals, described in detail in Section~\ref{sec:injections}, we compare the fractional errors in PyCBC Live's estimation of the redshifted (`detector frame') chirp mass $\mcdet = (1+z)\mchirp$ and in the mass ratio $q\equiv m_2/m_1$. The comparison is shown in Fig.~\ref{fig:mchirp_rel_err}, where errors in the search recovered $\mcdet$ and $q$ are presented for several ranges of $\mcdet$.
For low chirp mass binaries, considered more interesting for possible follow-up, the relative error in $\mchirp$ is around two orders of magnitude smaller than the relative error in $q$. For high chirp mass binaries the difference between the uncertainties decreases to under one order of magnitude.  These trends roughly match expectations from overlap calculations in~\cite{Biscoveanu:2019ugx}. 

The difference in accuracy, along with the mentioned bias caused by redshift, suggest an improvement in the source classification for PyCBC Live: the proposed new method is described in the next section.

\section{Methods}
\label{sec:methods}

In this section we will describe a method based on the search pipeline outputs to estimate the probability of a candidate event to belong to one of the following four astrophysical categories: \ac{BNS}, \ac{NSBH}, \ac{BBH} or MassGap. For this classification, we consider every object with a mass below 3\,\msun\ as a \ac{NS}, every object with a mass above 5\,\msun\ as a \ac{BH}, and every candidate event with at least one component mass between these limits as a MassGap.
\begin{figure}[b!]
    \centering
    \includegraphics[width=\columnwidth]{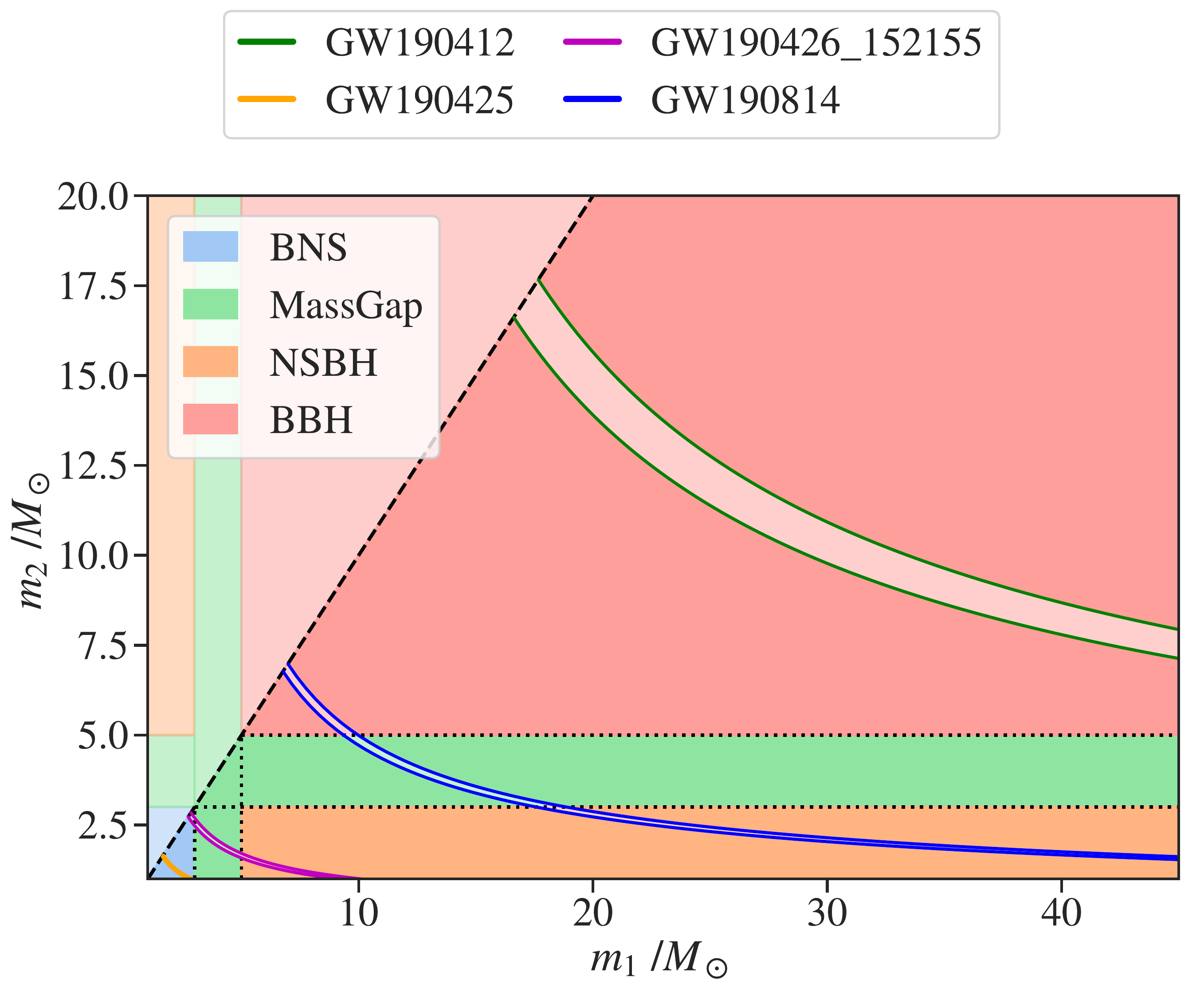}
    \caption{Chirp mass contours over the plane of component masses, $m_1$, $m_2$, for several O3 PyCBC Live triggers. Our (source frame) chirp mass estimates are: 
    GW190412 ($14.91\pm0.46\,\msun$), 
    GW190425 ($1.43\pm0.02\,\msun$), 
    GW190426\_151255 ($2.42\pm0.05\,\msun$), 
    GW190814 ($5.99\pm0.09\,\msun$)
    .}
    \label{fig:m1m2_contours}
\end{figure}

As stated in Section~\ref{sec:motivation}, the method uses the chirp mass~\eqref{eq:mchirp} recovered by the search pipeline as an input. Since this value is measured with high accuracy (up to a correction due to redshift, which we will discuss below), the component masses $m_1$ and $m_2$ are constrained to be close to a line of constant chirp mass over the mass plane. Assuming an uncertainty on the chirp mass, $\Delta \mchirp$, this line becomes a contour of $\mchirp \pm \Delta \mchirp$ over the mass plane; we discuss our uncertainty estimate $\Delta \mchirp$ in more detail below.  In Fig.~\ref{fig:m1m2_contours} we show some chirp mass contours corresponding to PyCBC Live triggers of events of \ac{O3a}.
\begin{figure*}[t!]
    \begin{minipage}{.49\textwidth}
        \centering
        \includegraphics[width=\linewidth]{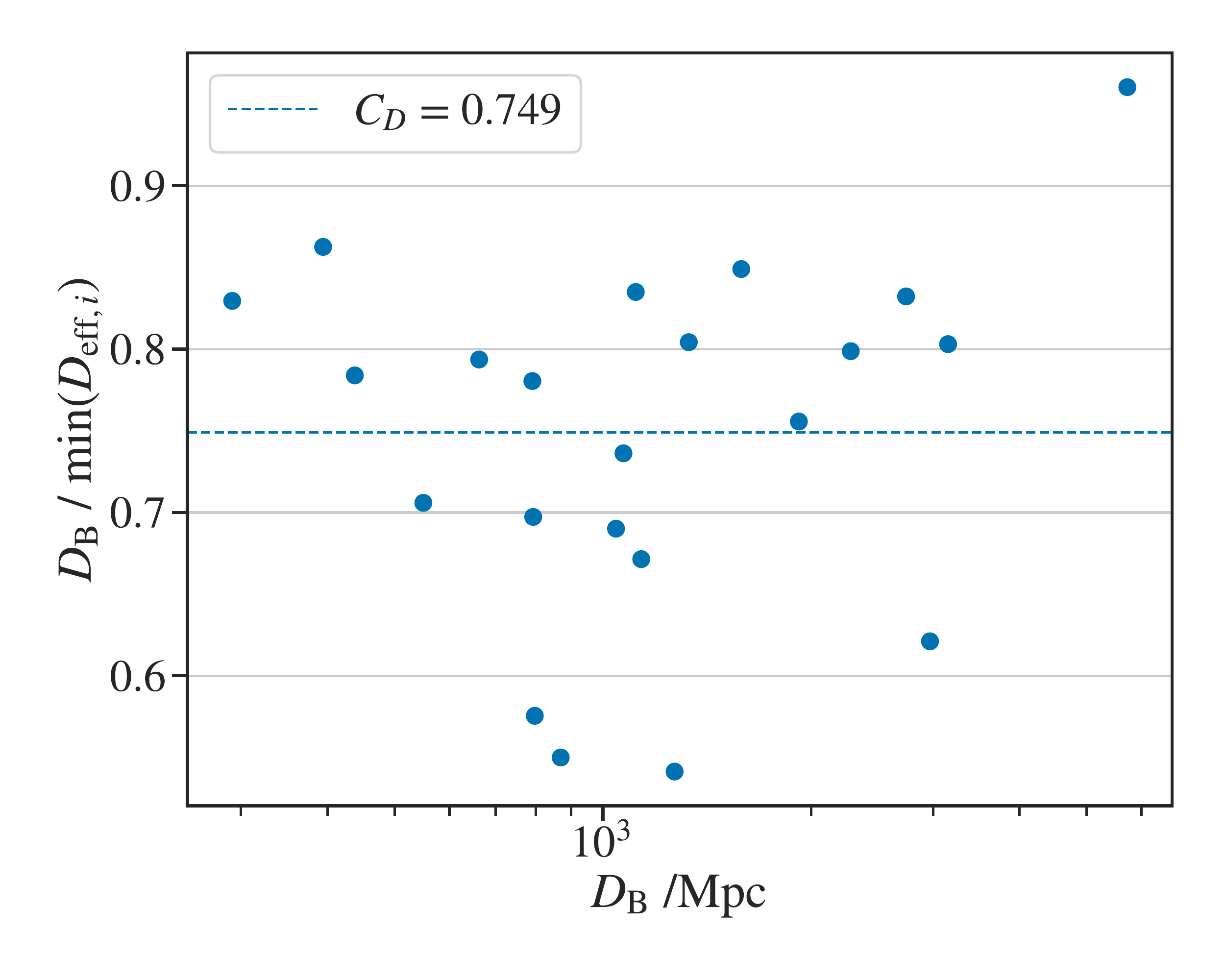}
        \label{fig:min_deff_coeff}
    \end{minipage}
    \begin{minipage}{.49\textwidth}
        \centering
         \includegraphics[width=\linewidth]{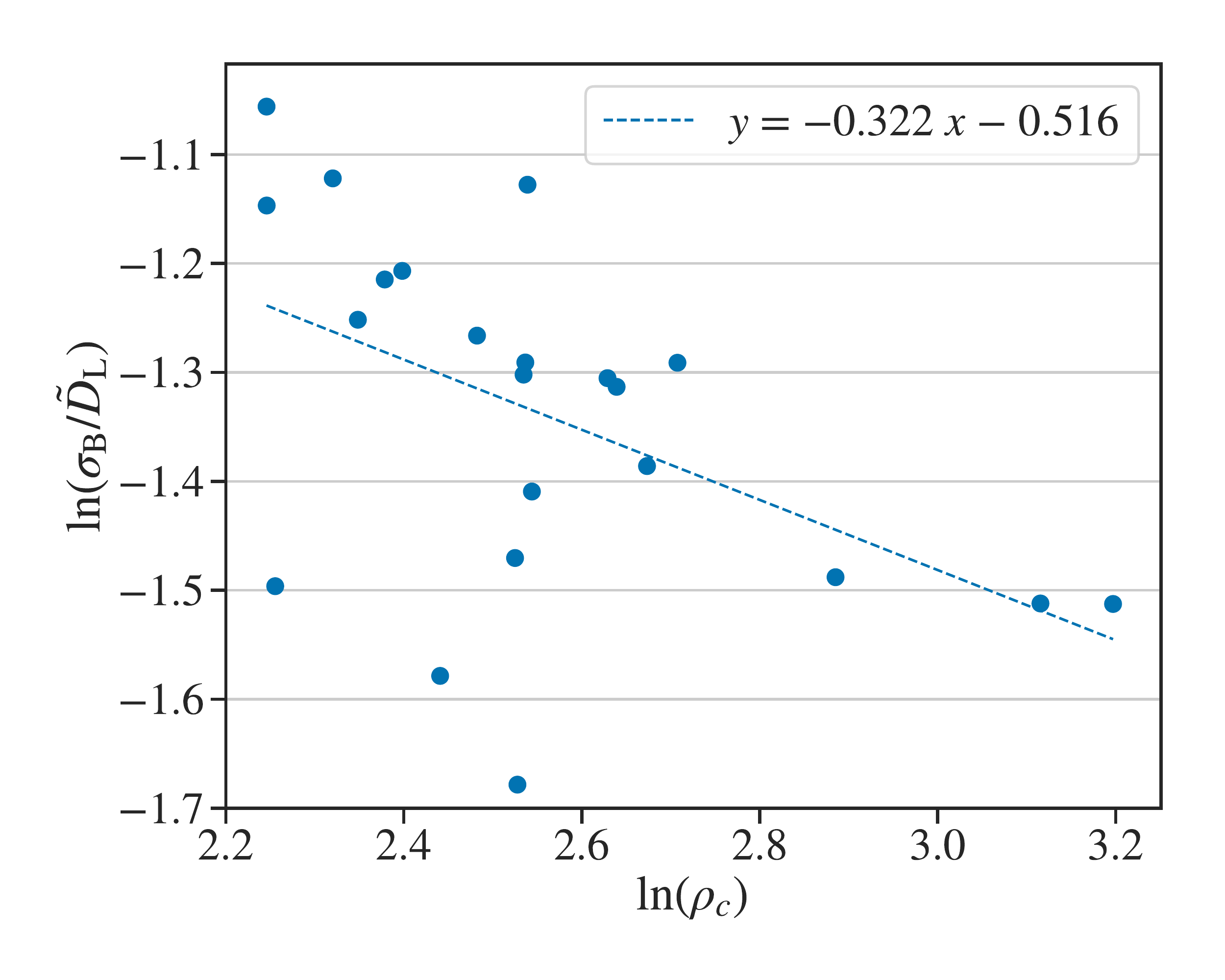}
        \label{fig:stdbay_dbay_snr_coeff}
    \end{minipage}
    \vspace{-0.5cm}
    \label{fig:fittings}
    \caption{Fitting plots for the luminosity distance $\dest$ (left) and its uncertainty $\sigmaest$ (right).
    }
\end{figure*}
The main idea of this estimation method is that the probabilities of the different categories are taken to be proportional to the area of each region of the plane that is inside the contour of chirp mass. The method thus assumes an uniform prior of candidate signals over the plane of component masses $m_1$, $m_2$. The output of this method for a candidate event is a list of probabilities $\{p_\mathrm{BNS}, p_\mathrm{MG}, p_\mathrm{NSBH}, p_\mathrm{BBH}\}$ summing to unity.

Component masses $m_1$, $m_2$, measured in the source frame, are constrained to the range $1 < m/\msun < 45$, which results in a chirp mass range $0.87 < \mchirp/\msun < 39.17$. The lower bound on the component masses is the lower limit on the template bank mass space adopted by the pipeline during \ac{O3a} \cite{DalCanton:2017ala}, and the upper bound is chosen based on \ac{BBH} population studies from the first and second observing runs (O1 and O2) \cite{LIGOScientific:2018jsj} and \ac{O3a} \cite{LIGOScientific:2020kqk}. Given these limits, every object with a chirp mass lower than 
0.87\,\msun\ will be considered a \ac{BNS}, and every object with a chirp mass greater that 
39.17\,\msun\ will be considered a \ac{BBH}.

However, the chirp mass value recovered by the pipeline is that of the matched filter templates: since the templates are designed to match the signal as it appears in the data stream, the template mass corresponds to the redshifted chirp mass, $\mcdet = (1+z)\mathcal{M}$. To recover the true chirp mass of the event we estimate the redshift by assuming some set of cosmological parameters to relate it to the luminosity distance $D_\mathrm{L}$.

One can deduce $D_\mathrm{L}$ from the amplitude of a \ac{CBC} signal at a given detector, which is inversely proportional to $D_\mathrm{L}$ but also depends on the masses and spins of the source and on its position and orientation relative to the detector.  
For low-latency analysis, the luminosity distance is estimated by the rapid Bayesian algorithm \texttt{BAYESTAR}~\cite{PhysRevD.93.024013} as a part of the sky localization process. This follow-up process starts after the candidate event trigger has been uploaded to \ac{GraceDB} and hence the search pipeline does not have access to \texttt{BAYESTAR} estimated luminosity distances. To overcome this problem we estimate the luminosity distances using a parameter called \textit{effective distance} $\deff$~\cite{Allen:2005fk}. 

The expected \ac{SNR} of a signal in a detector $i$, assuming that the template waveform is perfectly matched to the signal, is given by
\begin{equation}
\label{eq:deff_definition}
    \bar{\rho}_i = \frac{\sigma_i}{\deffi},
\end{equation}
where $\sigma_i$ is a (mass- and spin-dependent) measure of the detector sensitivity given by the expected $\ac{SNR}$ for a source at 1\,Mpc distance, directly overhead the detector and with zero inclination (angle between the direction to the observer and the orbital angular momentum of the binary)~\cite{Allen:2005fk}. The effective distance $\deffi$ is related to the luminosity distance $D_\mathrm{L}$ by geometrical factors as 
%
\begin{equation}
    \deffi = D_\mathrm{L}\left[F_{+,i}^2\left(\frac{1+\cos^2\iota}{2}\right)^2
    + F_{\times,i}^2 \cos^2\iota\right]^{-1/2},
    \label{eq:eff_dist}
\end{equation}
where $F_+$ and $F_\times$ are the antenna response functions for the incident signal, and $\iota$ is the inclination angle~\cite{Allen:2005fk}:  
note that $\deffi \geq D_\mathrm{L}$ for all detectors. 

Using Eq.~\eqref{eq:deff_definition}, the effective distance is estimated by PyCBC Live for each of the instruments present for a given event as
\begin{equation}
    \hat{D}_{\mathrm{eff},i} = \frac{\sigma_i}{\rho_i} \: \mathrm{Mpc},
    \label{eq:eff_dist_pycbc}
\end{equation}
where $\rho_i$ is the matched filter \ac{SNR}.
\begin{figure*}[htb]
    \centering
    \includegraphics[width=\textwidth]{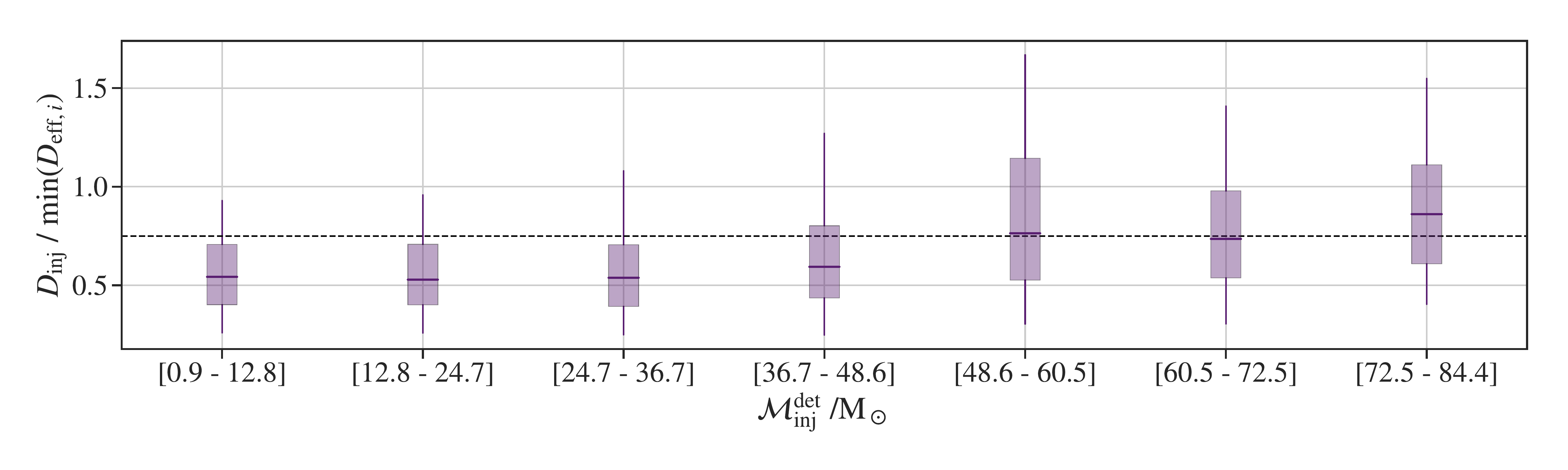}
    \caption{Boxplot showing the ratio between the simulated (true) distances and the minimum effective distances recovered by PyCBC Live for different detector-frame chirp mass intervals. Dotted line corresponds to a value of $\dinj/\min(\deffi)=0.749$, the fitting constant of Eq.~\eqref{eq:est_dist}.}
    \label{fig:check_dist_mc}
\end{figure*}

To estimate the luminosity distance using these estimated effective distances, we study the relation between them. 
We collect the $\hat{D}_{\mathrm{eff},i}$ values in each detector and the luminosity distance computed using \texttt{BAYESTAR}, $D_\mathrm{B}$, for PyCBC Live triggers from all the public candidate events from \ac{O3a}.  We do not expect that the effective distance estimated by any one pre-determined detector will be a good predictor of $D_\mathrm{B}$, so we choose for each candidate event the closest effective distance to $D_\mathrm{B}$, which is also the smallest one: $\min(\deffi)$. We then estimate luminosity distances as
\begin{equation}
    \tilde{D}_\mathrm{L} = C_D \cdot \min(\hat{D}_{\mathrm{eff},i}),
    \label{eq:est_dist}
\end{equation}
where $C_D$ is a constant, for which we take the mean of the ratios $D_\mathrm{B}$/$\min(\deffi)$ of all selected triggers to obtain $C_D=0.749$.
This 
can be seen in the left plot of Fig.~\ref{fig:fittings}.
We also estimate the uncertainty on luminosity distance using the same selection of triggers from \ac{O3a}, by performing a log-linear regression of the ($1\sigma$) luminosity distance uncertainties computed using \texttt{BAYESTAR}, $\sigma_\mathrm{B}$, relative to the estimated luminosity distances of Eq.~\eqref{eq:est_dist}, against the network \ac{SNR} of the trigger, $\rho_\mathrm{c}$.\footnote{$\rho_c = \sqrt{\sum_i \rho_i^2}$, being $i$ each detector present in the event detection.}
This fit, shown in the right plot of Fig.~\ref{fig:fittings}, gives a estimate of the uncertainty 
\begin{equation}
    \tilde{\sigma}_{D_\mathrm{L}} = \mathrm{e}^{-0.516} \cdot \tilde{D}_\mathrm{L} \cdot \rho_\mathrm{c}^{-0.322}.
    \label{eq:est_sigma_dist}
\end{equation}
From the estimated luminosity distance and relative uncertainty, we derive the redshift $\tilde{z}$ and redshift uncertainty $\tilde{\sigma}_z$ estimates assuming a standard flat $\mathrm{\Lambda}$CDM cosmology, where
\begin{equation}
    D_\mathrm{L}(z)=\frac{c(1+z)}{H_0}\int_0^z\frac{dz'}{\Omega_m(1+z')^3+(1-\Omega_m)},
\end{equation}
with Hubble parameter $H_0=67.90\,$km\,s$^{-1}$\,Mpc$^{-1}$ and matter density parameter $\Omega_m=$ 0.3065~\cite{Planck:2015fie}. 

To finally estimate the source frame \mchirp\ uncertainty $\Delta\mchirp$, we combine the propagated redshift uncertainty $\tilde{\sigma}_z$ with a small nominal uncertainty of 1\% in the detector-frame chirp mass recovered by the search pipeline.  As shown in Fig.~\ref{fig:mchirp_rel_err}, this 1\% nominal error is a representative value for signals around the centre of the total mass range we consider.  Also, in most cases we expect the redshift uncertainty will be comparable to or larger than uncertainty in \mcdet.  For higher-mass signals, a highly accurate estimate of $\Delta\mchirp$ is not necessary as the classification will be dominated by \ac{BBH}; for low-mass signals, the 1\% nominal error is an overestimate, however we have verified that the effect on the resulting source probabilities is negligibly small.

\section{Check with simulated signals}
\label{sec:injections}
\begin{figure*}[hbt]
    \centering
    \includegraphics[width=\textwidth]{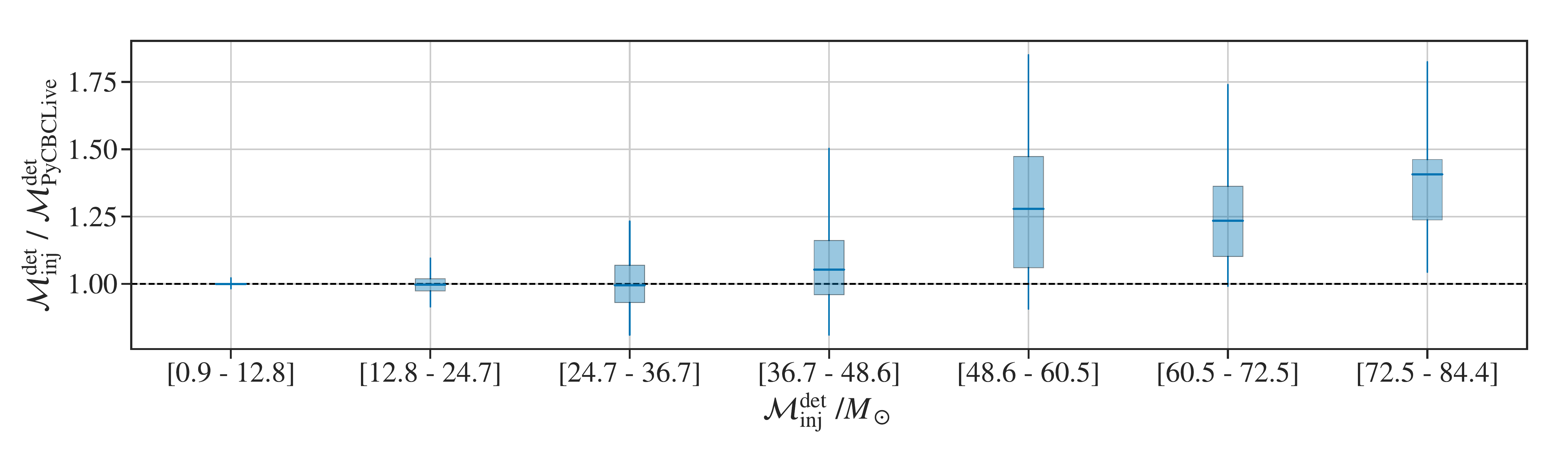}
    \caption{Boxplot showing the ratio between the simulated chirp masses and the chirp masses measured by PyCBC Live for different chirp mass intervals. All chirp masses are in the detector frame. Dotted line corresponds to a ratio equals to unity.}
    \label{fig:check_mchirp_bias}
\end{figure*}
In order to check the accuracy of the classification method and test the luminosity distance fitting constants used in Eq.~\eqref{eq:est_dist} and \eqref{eq:est_sigma_dist}), we use a population of simulated astrophysical signals injected into the \ac{O3a} data and recovered using PyCBC Live. The simulated population campaign was originally performed for testing the improvements made to PyCBC Live for \ac{O3}, and therefore is described in detail in \cite{DalCanton:2020vpm}. In summary, for the simulated binaries, \aclp{NS} have masses distributed uniformly between 1\,$\msun$ and 3\,$\msun$, and \aclp{BH} have masses distributed uniformly between
3\,$\msun$ and 97\,$\msun$. This masses are measured in the detector frame. 
All \ac{CBC} simulated sources are distributed uniformly in \textit{chirp distance} \cite{Allen:2005fk} between 5$-$300\,Mpc.

\subsection{Accuracy of distance estimation}

First, we present a test on the reliability of the luminosity distance estimation of Eq.~\eqref{eq:est_dist}. In Fig.~\ref{fig:check_dist_mc} we show a boxplot of the ratio of the simulated luminosity distances, $\dinj$, and the minimum effective distances recovered by PyCBC Live, $\min(\deffi)$ for detector-frame chirp mass intervals between 0.9-84.4\,$\msun$. The dotted line in the plot represents the fit obtained in the previous section, $C_D = 0.749$. 
The plot shows a deviation from the fitting constant for low chirp mass and very high chirp mass binaries. Up to $\sim$50\,$\msun$, this ratio is overestimated; 
conversely, for chirp mass values over $\sim$70\,$\msun$, the fitting constant is underestimated.  
As expected, the fit is more accurate for chirp masses toward the centre of the range, since it was estimated using \ac{O3a} events which mostly fall into this region.

A possible reason for this bias may be that the search template sensitivity $\sigma_i$, used in Eq.~\eqref{eq:eff_dist_pycbc} to estimate the source distance, is biased with respect to the sensitivity for a template corresponding to the true source parameters.  In Fig.~\ref{fig:check_mchirp_bias}, we present a boxplot of the ratio between the injected (redshifted) chirp masses and the chirp masses recovered by PyCBC Live, for chirp mass intervals between 0.9-84.4\,$\msun$: for low chirp mass signals the search shows an unbiased recovery, however for signals with $\mchirp(1+z)\gtrsim 50\,\msun$ there is a clear bias towards underestimating injection chirp masses, which may result from the limited extent of the template bank.  Since the sensitivity $\sigma_i$ is largely determined by the template chirp mass, this recovery bias will contribute to a bias in the estimation of source distances.

For low-chirp-mass systems, which are of most interest for potential \ac{EM} follow-up, we may thus obtain a distance estimate that is not affected by this bias, at the cost of a 
higher bias for high-mass \ac{BBH} signals. 
For this, we reproduce the linear fit of the previous section taking the mean of the ratio $\dinj/\min(\deffi)$ only for simulated signals with $\mcdet_\mathrm{inj}<40\,\msun$, giving a revised value of $C_D = 0.576$.  However, in the remainder of this work our results use the value $C_D = 0.749$ from a fit of \ac{O3a} public alert events. 

\subsection{Accuracy of classification method}

In order to check the accuracy of the chirp-mass based classification, we impose an additional constraint on the simulated signal parameters. Given the component mass limits described in Section~\ref{sec:methods}, asymmetric high-mass \ac{NSBH} systems outside these limits are not representative of the accuracy of the method. Therefore, we restrict the black hole components of simulated \ac{NSBH} events to be below 50\,$\msun$.

We present two different visualizations of the performance of the method: a confusion matrix and a \ac{KDE} of predicted probabilities.

\begin{figure}[h]
    \centering
    \includegraphics[width=\columnwidth]{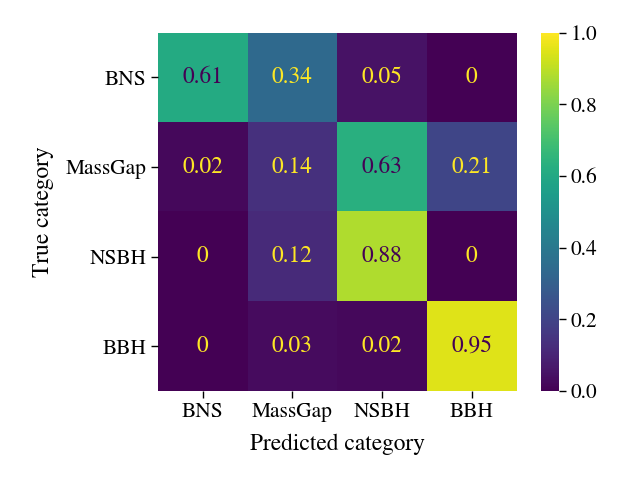}
    \caption{Confusion matrix comparing the true categories of the simulated signals versus the category found with highest probability in each event.}
    \label{fig:inj_CM}
\end{figure}

The confusion matrix, shown in Fig.~\ref{fig:inj_CM}, compares the category of the simulated signals (\textit{true category}), defined by their given source masses, against the category found with the highest probability by the classification method (\textit{predicted category}). 

\ac{BBH} and \ac{NSBH} simulated events are assigned with highest probability to their correct categories in a large majority of cases: 95\% and 88\% respectively. \ac{BNS} signals have a 61\% probability of being classified correctly, but a notable 34\% of cases are found more likely to be MassGap sources. Simulations of MassGap sources are mostly classified as \ac{NSBH}, with a 63\% probability, followed by 
\ac{BBH} and 
MassGap. This bias between \ac{NSBH} and MassGap categories can be justified considering the high uncertainty on rates and masses of both \ac{CBC} sources, and implies a conservative approach of the classification method: it will suggest a follow-up for possible \ac{EM} counterparts of \ac{NSBH} signals even if the source is a MassGap binary.

\begin{figure}[bht]
    \centering
    \includegraphics[width=\columnwidth]{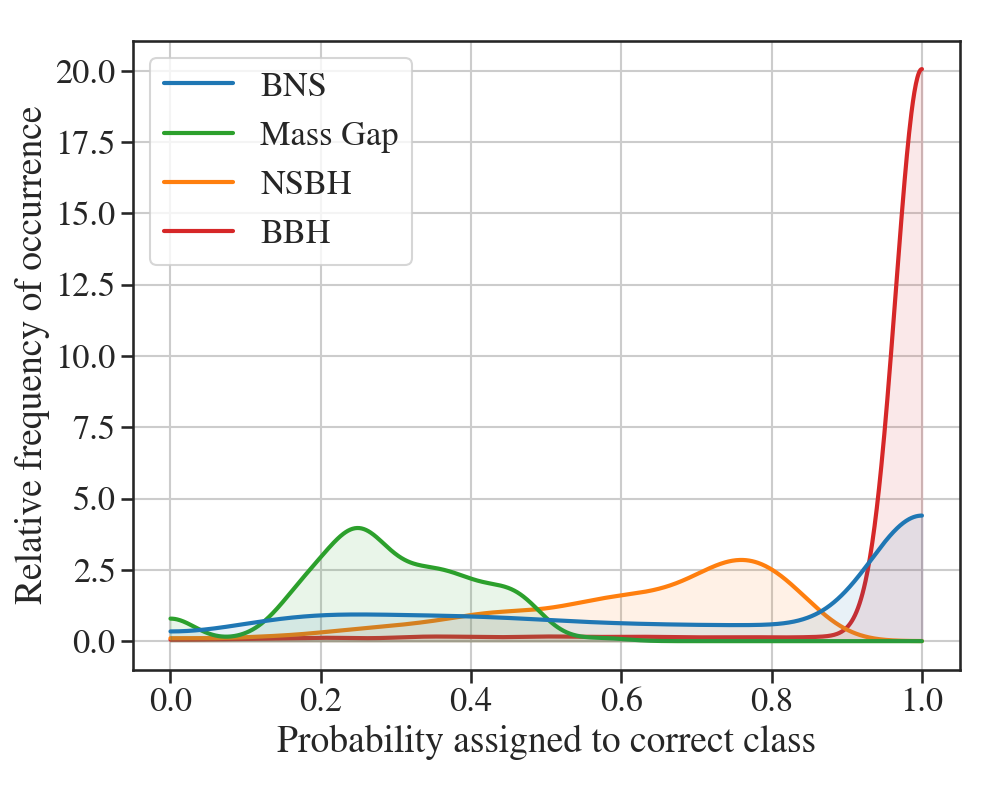}
    \caption{Density distribution of the probabilities estimated by our method for the correct \ac{CBC} category of the simulated population. Correct categories are determined from the given source component masses.}
    \label{fig:inj_KDE}
\end{figure}

The \ac{KDE} plot, presented in Fig.~\ref{fig:inj_KDE}, shows the distribution of the estimated probabilities that our method assigns to the correct \ac{CBC} category of each simulated signal.  The correct category is determined from the given source component masses of the signal, according to the mass boundaries described in Section~\ref{sec:methods}.

\ac{BNS} and \ac{BBH} simulated signals are assigned correctly with probabilities over 80\% for a large or very large majority of cases. These high probabilities are expected, due to the location of both classes in the plane of component masses $m_1$, $m_2$: chirp mass contours over \ac{BNS} or \ac{BBH} areas do not necessarily overlap other target regions.
This is not the case of \ac{NSBH} or MassGap simulated events, whose chirp mass contours always overlap other source areas, and this lowers the probabilities assigned to the correct category.
The majority of \ac{NSBH} simulations are assigned to the correct class with probabilities $p_\mathrm{NSBH}$ over 50\%, but the narrowness of the MassGap region causes the $p_\mathrm{MG}$ probabilities to fall under the 50\% for the majority of cases.


\subsection{Removal of MassGap category}
\label{subsec:injections-nomg}

The MassGap category was included in the \ac{CBC} source classification at the beginning of \ac{O3}, motivated by the possible existence of a limit on the minimum mass of a \acl{BH} around 5\,\msun. At the time of writing, heading towards the start of \ac{O4}, there are suggestions for changing the status of MassGap from a source class to a source property, along with the already existing \textit{HasNS} and \textit{HasRemnant}~\cite{EMFollowUserGuide}. 
In view of this potential change, we present the same visualization plots for the chirp-mass based classification of the mentioned population of simulated signals in the remaining \ac{BNS}, \ac{NSBH} and \ac{BBH} categories: in this case, every component with mass greater than 3\,\msun\ is considered a \ac{BH}. 
\begin{figure}[b!]
    \centering
    \includegraphics[width=\columnwidth]{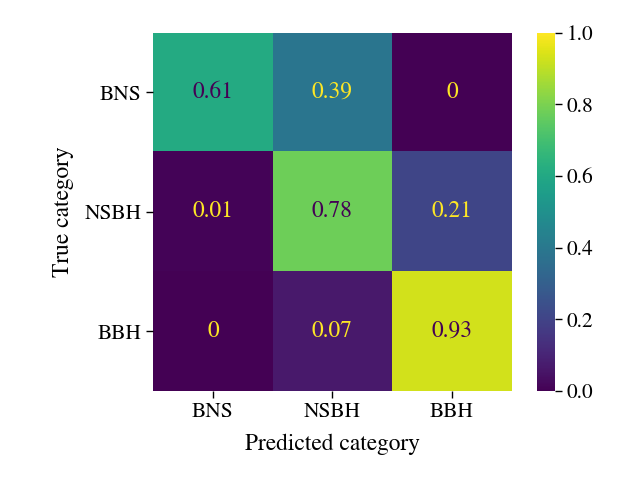}
    \caption{Confusion matrix comparing the true categories of the simulated signals versus the category found with highest probability by our method for \ac{BNS}, \ac{NSBH} and \ac{BBH} categories.}
    \label{fig:inj_CM_noMG}
\end{figure}

The confusion matrix of Fig.~\ref{fig:inj_CM_noMG} shows a slight change in the probability of a \ac{BBH} to be assigned correctly, from 95\% to 93\%, where the remaining 7\% will be classified as \ac{NSBH}. In the case of \ac{NSBH} simulated signals, the rate of correct assignations changes from 88\% to 78\%, and 21\% of cases are now classified as most likely to be \ac{BBH}. Lastly, there is no change in the 61\% of \ac{BNS} simulations correctly assigned, and the remaining cases are found as most likely to be \ac{NSBH}, meaning that an \ac{EM} follow-up would be always recommended for this range of masses.

\begin{figure}[htbp]
    \centering
    \includegraphics[width=\columnwidth]{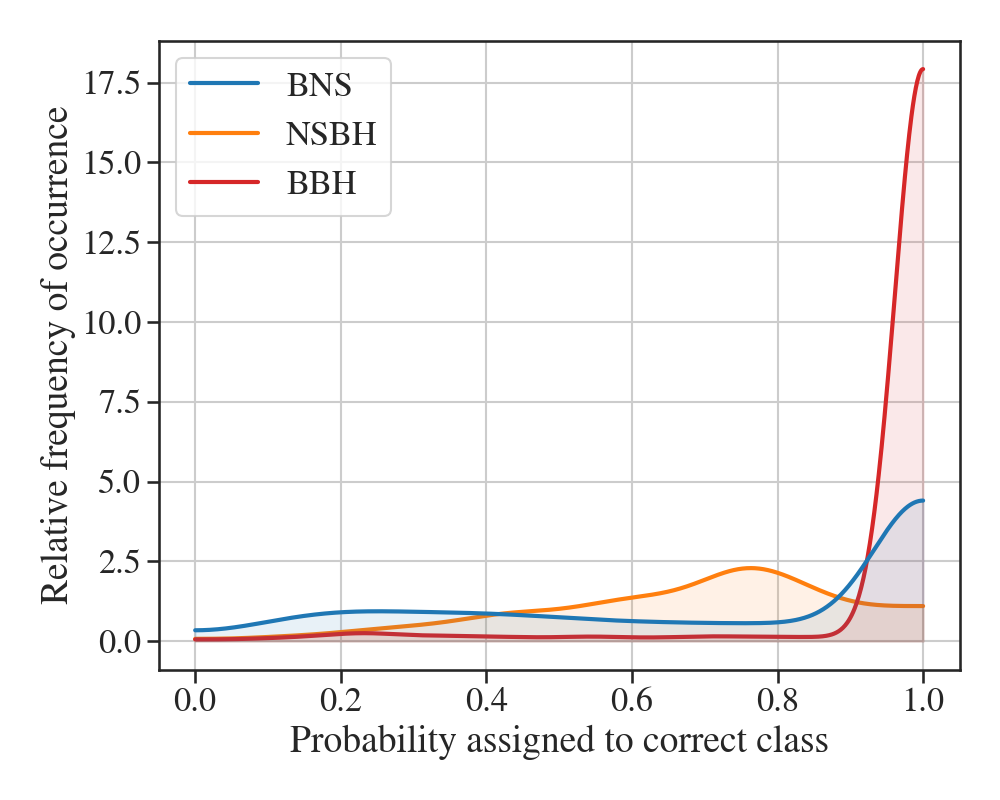}
    \caption{Density distribution of the probabilities estimated by our method for the correct \ac{CBC} category of the simulated population, among \ac{BNS}, \ac{NSBH} and \ac{BBH}. Correct categories are determined from the given source component masses.}
    \label{fig:inj_KDE_noMG}
\end{figure}

The \ac{KDE} plot, presented in Fig.~\ref{fig:inj_KDE_noMG}, shows a main change in the distribution of \ac{NSBH} probabilities, where a notable amount of cases are now assigned very high correct class probabilities. This behaviour is consistent with the decrease of the overlap between the \ac{NSBH} area with the other source class regions.

\section{Results for O3 confirmed events}
\label{sec:o3results}

In this section we present the probabilities obtained applying the chirp-mass based classification method to 
26 events detected during \ac{O3a}, and 19 during \ac{O3b}. The selected events follow these criteria: they were candidate events disseminated in the public \ac{GCN} notices and circulars during \ac{O3}, and they pass the publication thresholds of \ac{GWTC-2}~\cite{LIGOScientific:2020ibl} and \ac{GWTC-3}~\cite{LIGOScientific:2021djp}, respectively.

We want to compare the accuracy of our method against the classification sent in the public low-latency \ac{GCN} Circulars and the detailed \ac{PE} results obtained afterwards (offline) for \ac{GWTC-2} and \ac{GWTC-3}. 
The results are presented in Table~\ref{tab:general-events}, where the second column gives the probabilities obtained by the method described in this paper. 
These results are compared with the probabilities distributed in the \ac{GCN} Circulars \cite{LIGOPublicGCNs}, in the third column, and with the probabilities obtained by the \ac{PE} techniques of \ac{GWTC-2} and \ac{GWTC-3}, in the fourth column. The source chirp mass computed with \ac{PE} is also shown as context for further discussion. The section is organized as follows: in the three first subsections we describe in more detail how the probability estimates were obtained for each of the methods we compare, and the last subsection is devoted to an extended discussion of this comparison.
\begin{table*}[tbhp]
    \centering
    \resizebox{0.9\textwidth}{!}{\input{table_results}}
    \caption{
    Source classification percentage probabilities for confirmed \ac{O3a} events identified in low-latency alerts.
    chirp-mass based probabilities, in the second column, are estimated using search template chirp masses and estimates of the distance to the source. PyCBC Live distance estimates are available for all events, except the ones with a dagger ($^\dagger$), where GstLAL triggers and \texttt{BAYESTAR} distance estimates are used.
    Probabilities issued as low-latency public alerts, 
    in the third column, are derived from the preferred search trigger in \ac{GraceDB}. Bold-faced events correspond to PyCBC Live preferred triggers, events in italics correspond to MBTAOnline preferred triggers, typewriter-font corresponds to a SPIIR preferred trigger, and the rest correspond to GstLAL preferred triggers.
    If two sets of \ac{GCN} probabilities are shown, the second set is 
    a later update based on posterior support from parameter estimation.
    %
    To compute \ac{PE} probabilities, in the fourth column, we use samples reweighted to a prior with uniform merger rate in the co-moving frame.
    The source chirp mass computed from \ac{PE} is also shown.
    }
    \label{tab:general-events}
\end{table*}

\subsection{Chirp-Mass based probabilities}

We apply the chirp-mass based method to a total of 45 online triggers belonging to public candidate events that were submitted to \ac{GraceDB} near the time of the event. The same candidate event may be reported to \ac{GraceDB} by different analysis pipelines, so a trigger choice has to be made in order to perform the analysis. The trigger selection is motivated as follows: as we plan to implement this method in PyCBC Live, we use the data of the online trigger of this pipeline, if available; if no PyCBC Live trigger is available, we take the information of the \textit{preferred trigger} of the event. The criteria for choosing the \textit{preferred trigger} can be found in the LIGO/Virgo Alerts User Guide~\cite{EMFollowUserGuide}. Trigger data is not publicly available, but it can be computed analyzing the Data Releases of \ac{GWTC-2}~\cite{GWTC-2:DataRelease} and \ac{GWTC-3}~\cite{GWTC-3:DataRelease} with the corresponding online pipeline codes.

\paragraph{O3a events}
For 21 of the 26 candidate events selected, a PyCBC Live trigger is available in the \ac{GraceDB} database. To apply the chirp-mass based method, we extract from them their chirp mass, their effective distances and their \ac{SNR}. The candidate event GW190425 is a special case, since it was detected by PyCBC Live as a LIGO Livingston single-detector trigger. During \ac{O3}, the single-detector significance calculation method \cite{DalCanton:2020vpm} was not yet implemented into the PyCBC Live pipeline, and thereby the trigger was not reported in \ac{GraceDB}. However, since it was detected by PyCBC Live in low latency, we use its single-trigger parameters to compute the chirp-mass based probabilities. 

The 4 remaining \ac{O3a} candidate events without a PyCBC Live trigger are GW190513\_205428, GW190630\_185205, GW190728\_064510 and GW190915\_235702. All of them have a \textit{preferred trigger} from the GstLAL online search. Since this pipeline does not compute effective distances, we directly use the preliminary luminosity distances computed with \texttt{BAYESTAR} to apply the chirp-mass based method, besides the chirp mass and the \ac{SNR} values.


\paragraph{O3b events}
In this case, 14 candidate events out of the 19 selected have a PyCBC Live trigger available in \ac{GraceDB}. As in the \ac{O3a} events, we use their chirp mass, their effective distances and their \ac{SNR} to compute the chirp-mass based probabilities.

The candidate GW200105\_162426 was detected by PyCBC Live as a LIGO Livingston single-detector trigger, and thereby was not reported in \ac{GraceDB}. Even so, we use its single-trigger parameters to compute the chirp-mass based probabilities. 

The 5 candidate events of \ac{O3b} without a PyCBC Live trigger are GW191204\_171526, GW191222\_033537, GW200112\_155838, GW200302\_015811 and GW200316\_215756. All of them have a \textit{preferred trigger} from the GstLAL online search, thus we compute the chirp-mass based probabilities using their \texttt{BAYESTAR} preliminary luminosity distances, chirp masses and \acp{SNR}.

\subsection{\ac{GCN} Alert probabilities}

The \ac{GCN} Notices and Circulars always include a probability of the source to be of terrestrial origin.  In order to make an effective comparison with the chirp-mass based method, we rescaled the \ac{GCN} Circular probabilities to the total probability of astrophysical origin. 

The set of probabilities sent on initial \ac{GCN} Circulars is computed within the LIGO/Virgo rapid alert infrastructure, where the probability estimation method depends on the pipeline that reported the \textit{preferred trigger} of the candidate event.


Some events in Table~\ref{tab:general-events} present two sets of \ac{GCN} Circulars probabilities. Second sets correspond to a later update based on posterior support from \ac{PE}. For GW190426\_152155, an initial Circular~\cite{GCN24237} was sent in low latency, giving the candidate event a rescaled 57\% \ac{BNS} probability, but a Source Classification update~\cite{GCN24411} was sent 10 days later assigning the event a 60\% probability of \ac{NSBH} origin. For GW190728\_064510, the low-latency Circular~\cite{GCN25187} reports a 52\% probability of being a MassGap, but a few hours later an update on Source Classification~\cite{GCN25208} was made public, including a \ac{PE}-based classification that gives the event a 95\% \ac{BBH} probability. Lastly, for GW190814, the source classification sent in the initial Circular~\cite{GCN25324} was based on point mass estimates, assigning an estimate of 100\% MassGap probability to the candidate event. A few hours later, an update on Source Classification~\cite{GCN25333} was available, where the \ac{PE}-based classification gives a $>99\%$ \ac{NSBH} probability. Lastly, for GW191216\_213338, an initial Circular~\cite{GCN26454} was sent in low latency, giving the candidate a 100\% probability of MassGap origin. A week later, a Source Classification update~\cite{GCN26570} was made public, including a $>99\%$ \ac{BBH} probability.

The probabilities sent in Notices and Circulars are limited in precision to 1\% and in some cases are stated as an upper or lower limit.  We use numerical values obtained from \texttt{p\_astro.json} data files available at \ac{GraceDB}.

\subsection{Probabilities from \ac{GWTC-2} and \ac{GWTC-3} Parameter Estimation}

We may compare the chirp-mass based probabilities with the \ac{PE} results computed for the catalogs using samples published in the Data Releases of \ac{GWTC-2}~\cite{GWTC-2:DataRelease} and \ac{GWTC-3}~\cite{GWTC-3:DataRelease}.  Here we use samples reweighted to have a luminosity distance prior that corresponds to a uniform merger rate in the co-moving frame of the source, which are also used for \ac{PE} results in \ac{GWTC-2} and \ac{GWTC-3}.
The probabilities for the four \ac{CBC} categories are computed by distributing the samples in the different regions of the component mass plane, counting how many samples fall in each area and dividing them between the total number of samples: we use \texttt{PEPredicates} code\footnote{\url{https://pypi.org/project/pepredicates/}} accessed via the \texttt{PESummary} package~\cite{Hoy:2020vys} to carry out the calculation. 

For some events, for instance GW190426\_152155, a fraction of samples have component mass under the $1\msun$ limit usually taken as the minimum \ac{NS} mass.
These samples are not assigned to any astrophysical class, thus the sum of probabilities is not 100\% in such cases.  In contrast, our method generally assumes zero probability of a source having component mass $<1\msun$, thus our probabilities do sum to 100\%. 


\subsection{Discussion of results}

Finally, in this section we discuss the main features of the method for confirmed \ac{GW} events identified in low latency during \ac{O3}, and compare them with the classifications issued via \ac{GCN} and with those resulting from detailed offline \ac{PE}.
\begin{figure}[t]
    \centering
    \includegraphics[width=\columnwidth]{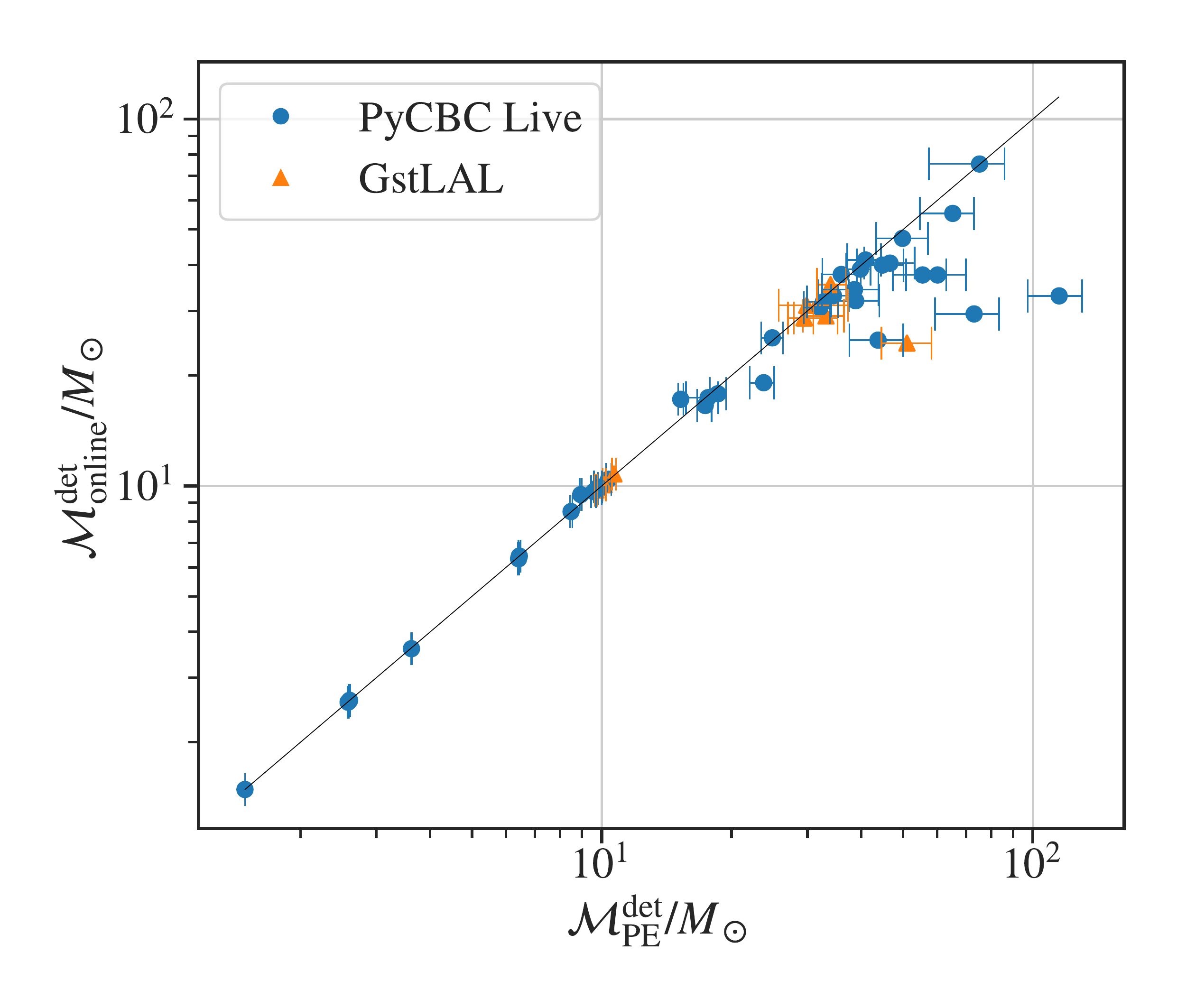}
    \caption{Scatter plot of the $\mcdet$ estimated by the low-latency pipelines, against the $\mcdet$ (median and 90\% credible interval) from \ac{PE} samples for the 45 \ac{O3} events considered in this paper. Chirp masses of PyCBC Live triggers are plotted if available;
    if not available, the trigger chirp mass of the preferred event is plotted (here, all preferred events are from GstLAL).
    } 
    \label{fig:mc_online_vs_PE}
\end{figure}
\begin{table*}[htb]
    \centering
    \input{table_results_noMG}
    \caption{
    Source probabilities for O3 events without the MassGap category (in \%). Only events with a probability of \ac{BBH} less than 100\% are shown. PyCBC Live distance estimates are available for all events, except the ones with a dagger ($^\dagger$), where GstLAL triggers and \texttt{BAYESTAR} distance estimates are used.}
    \label{tab:no-MG}
\end{table*}
The events naturally fall into several groups according to the \ac{PE} chirp mass.  In Fig.~\ref{fig:mc_online_vs_PE} we compare the redshifted chirp masses from the low-latency searches and the \ac{PE} analysis: there is no apparent bias except for higher $\mchirp$ binaries, where in some cases the searches underestimate the value relative to \ac{PE}. This deviation is expected at least for the PyCBC Live pipeline, given the results obtained using the simulated signals described in Section~\ref{sec:injections} (see Fig.~\ref{fig:check_mchirp_bias}).


For a large number of high- or moderately high-mass \ac{BBH} signals, all methods give probabilities of 100\% to the \ac{BBH} class, a result expected even if the $\mchirp$ and distance values used for the estimate have relatively large errors. There are several signals with somewhat lower chirp mass than the majority of \ac{BBH}, in the approximate range of $7-9$\,$\msun$, for which our method gives significantly large probabilities to the \ac{MG} class in addition to \ac{BBH}. These signals are all classified as \ac{BBH} with high probability by \ac{PE}, indicating that the data disfavour asymmetric masses in the \ac{MG} range, as opposed to our method which neglects information on mass ratio. However, our method still assigns high probability to the most likely true class (\ac{BBH}). The GCN probabilities for these events are either 100\% \ac{BBH} or 100\% \ac{MG} (for instance GW190930\_133541), thus in some cases assigning zero probability to the most likely origin. 

Two events, GW190814 and GW190924\_021846, have chirp mass consistent with \ac{BBH}, \ac{MG} and \ac{NSBH} origin.  They were assigned as 100\% \ac{MG} in GCN alerts, however detailed \ac{PE} indicates their probable true class as either certainly \ac{NSBH} (for GW190814), and conversely \ac{BBH} or \ac{MG} with nearly equal probability (GW190924\_021846).  Our method assigns \ac{NSBH} origin the highest probability for these events, with \ac{MG} the next most likely and small nonzero $p_\mathrm{BBH}$.  Thus in one case we correctly identify the most likely class, and in the other we at least do not assign zero probability to the correct class. 

Three further events have chirp masses inconsistent with \ac{BBH}, but allow \ac{MG} and \ac{NSBH} origin: the marginal candidate GW190426\_152155 and the first \ac{NSBH} detections, GW200105\_162426 and GW200115\_042309~\cite{LIGOScientific:2021qlt}.  All three are most likely \ac{NSBH} according to \ac{PE} analysis, with nonzero \ac{MG} probabilities for the first and last.  GCN alerts issued in low latency correctly identify the most likely class only for GW200105\_162426, whereas our method assigns all three as \ac{NSBH} with over 50\% probability (though also with significant nonzero $p_\mathrm{MG}$). 

Finally, only the event GW190425~\cite{LIGOScientific:2020aai} has a chirp mass consistent with \ac{BNS} origin. It was assigned a 100\% $p_\mathrm{BNS}$ by both our method and the low-latency \ac{GCN} alert, in agreement with the \ac{PE} result.
\begin{table*}[ht]
    \centering
    \resizebox{0.9\textwidth}{!}{\input{table_results_noz}}
    \caption{
    Source probabilities for O3 events, comparing the results with and without the redshift correction (in \%). Only events with a probability of \ac{BBH} less than 100\% are shown. PyCBC Live distance estimates are available for all events, except the ones with a dagger ($^\dagger$), where GstLAL triggers are used.  We also show our redshift estimates from search pipeline triggers, with a $1\sigma$ uncertainty estimate, and compare with the \ac{PE} values (median and symmetric $90\%$ interval).}
    \label{tab:no-redshift}
\end{table*}
\subsubsection{Results without MassGap category}
As previously discussed in Section~\ref{subsec:injections-nomg}, there are suggestions for removing the status of MassGap as a source class category. Therefore, we present the chirp-mass based probabilities of the 45 selected events to be consistent with a \ac{BNS}, \ac{NSBH} or \ac{BBH} origin. In this case, we only compare these probabilities against the ones obtained using the \ac{PE} samples. Results are presented in Table~\ref{tab:no-MG}. 

For events with $\mchirp > 10\,\msun$, there is no change with respect to the previous results: all were assigned a 100\% $p_\mathrm{BBH}$ by our method and by \ac{PE} analysis. Therefore, they are not shown in Table~\ref{tab:no-MG}. For lower chirp mass binaries with a high \ac{PE} probability of being \ac{BBH}, with $\mchirp$ between $7-9$\,$\msun$, the previous probability of \ac{MG} origin is transferred into the \ac{BBH} region, thus the \ac{BBH} probability now matches the \ac{PE} results.
For two events with a chirp mass consistent with \ac{MG}, \ac{NSBH} and \ac{BBH} origin (GW190814 and GW190924\_021846), their $p_\mathrm{MG}$ is assigned to the \ac{BBH} probablity. For both events, our method continues to find \ac{NSBH} as the most likely source class, followed closely by \ac{BBH}.
From the \ac{PE} analysis GW190814 is still assigned a 100\% probability of \ac{NSBH} origin, and GW190924\_021846 now has very high $p_\mathrm{BBH}$. 

Regarding the three events inconsistent with \ac{BBH} origin, but with some probability of being \ac{MG} or \ac{NSBH}, there are differences depending on whether they are consistent with \ac{BNS} or not. For GW190426\_152155, inconsistent with \ac{BNS} origin, its \ac{MG} probability is converted to \ac{BBH} probability. For GW200105\_162426 and GW200115\_042309, that have some probability of \ac{BNS} origin, their $p_\mathrm{MG}$ is allocated to the \ac{NSBH} class. Now, the three events are most likely \ac{NSBH} according to both our method and the \ac{PE} analysis.

The single event highly consistent with \ac{BNS} origin, GW190425, is as before assigned $p_\mathrm{BNS} \simeq 100\%$ both by our method and by \ac{PE}.

In general, the removal of the MassGap category results in our method being more consistent with the results from the \ac{PE} analysis.

\subsubsection{Effect of redshift correction}
Lastly, we discuss how the correction to the chirp mass for the bias caused by the redshift affects the distribution of source probabilities. We compare the probabilities obtained using our chirp-mass based method with and without this correction for the selected 45 events of \ac{O3}. Results are presented in Table~\ref{tab:no-redshift}, where only the events with $p_\mathrm{BBH}<100\%$ are shown; the table also compares our redshift estimates to the \ac{PE} values published in \ac{GWTC-2} and \ac{GWTC-3}.
In general, as ignoring the redshift bias overestimates the value of the chirp mass of the event, the probabilities are shifted towards the values that more massive systems would have. 
Also, the shifts in the assigned probabilities are of order $100\% \times z$. 

This comparison is done using the fit of Eq.~\eqref{eq:est_dist} with $C_D = 0.749$ as obtained from O3a events.  With the slightly smaller value $C_D = 0.576$, our estimated probabilities would be shifted slightly towards the values obtained without any redshift correction.  



Although the detailed pattern of changes in probability including the \ac{MG} class is complicated, we see that for instance for GW190425, omitting the redshift correction leads to a small but nonzero $p_\mathrm{MG}$.  At the other extreme, for events with a source chirp mass between $7-9\,\msun$, 
their $p_\mathrm{NSBH}$ falls to $\sim 0\%$, whereas their $p_\mathrm{BBH}$ grows significantly.  While this result happens to lead to estimated probabilities close to \ac{PE}, this outcome is due to an accidental combination of different biases.

We note that our estimated redshifts are almost all within the PE $90\%$ range, thus we may expect to able to obtain an unbiased redshift estimate (albeit with possibly large statistical errors).

\section{Conclusions}
\label{sec:conclusion}
In this paper we present a new method to directly estimate \ac{CBC} source distances and source classifications from the output of the PyCBC Live online search pipeline, while adding less than 50\,ms of alert latency. 
These classifications are obtained using the template chirp mass point estimates of the search, correcting for the bias caused by cosmological redshift. 
We applied the method to confirmed \ac{GW} events identified by LIGO-Virgo in low latency during \ac{O3}, as well as to a large population of simulated astrophysical signals.
For high and low chirp mass binary systems, classifications tend to be accurate, assigning high or very high probabilities to the correct class. On the other hand, for sources with intermediate chirp masses, the method always gives some probability to the correct category, but also significant nonzero probabilities to other categories. This behaviour is expected, due to systematic errors and statistical measurement uncertainty in the binary mass ratio, which our method does not completely account for. 
Estimates of source distances have a large statistical uncertainty, but we show that we can produce an unbiased estimate at least for systems with chirp masses up to 40\,$\msun$. 
For more massive systems, chirp masses and distances show a bias in recovery, possibly due to a limitation of the template bank of the search. 

Although this method was not deployed in \ac{O3}, in part due to the earlier than expected end of the run, it will be available for the next LVK observing run \ac{O4}.  As noted above, to obtain more accurate classification probabilities we must include information on the binary mass ratio.  To obtain an unbiased estimate we will also need to account for component spins; thus, future extensions of this method will consider a higher dimensional parameter space, as done in~\cite{Ohme:2013nsa}.

Going beyond the estimation of binary component properties, it is of interest for follow-up observations to predict the possibility of \ac{EM} emission from the binary based on those properties, as investigated in~\cite{Stachie:2021noh}.  Though, even knowledge of the (source frame) chirp mass alone, which we are able to estimate with high accuracy, is sufficient to predict many features of a \ac{BNS} merger and its subsequent evolution~\cite{Margalit:2019dpi}.

\section{Acknowledgments}
This work has received financial support from Xunta de Galicia (Centro singular de investigación de Galicia accreditation 2019-2022), by European Union ERDF, and by the “María de Maeztu” Units of Excellence program MDM-2016-0692 and the Spanish Research State Agency. VVO thankfully acknowledges support from the Ministry of Science, Innovation and Universities (MICIU) of Spain (FPI grant PRE2018-085436).
The authors would like to thank Gareth S. Cabourn Davies for his useful comments on an earlier version, and Tito DalCanton, Juan Calderón Bustillo and Bhooshan Gadre for discussions and technical advice.
The authors are grateful for computational resources provided by the LIGO Laboratory and supported by National Science Foundation Grants PHY-0757058 and PHY-0823459. Some of the results in this paper have been derived using the ``pesummary`` package~\cite{Hoy:2020vys}. This research has also made use of data or software obtained from the Gravitational Wave Open Science Center (gw-openscience.org), a service of LIGO Laboratory, the LIGO Scientific Collaboration, the Virgo Collaboration, and KAGRA. LIGO Laboratory and Advanced LIGO are funded by the United States National Science Foundation (NSF) as well as the Science and Technology Facilities Council (STFC) of the United Kingdom, the Max-Planck-Society (MPS), and the State of Niedersachsen/Germany for support of the construction of Advanced LIGO and construction and operation of the GEO600 detector. Additional support for Advanced LIGO was provided by the Australian Research Council. Virgo is funded, through the European Gravitational Observatory (EGO), by the French Centre National de Recherche Scientifique (CNRS), the Italian Istituto Nazionale di Fisica Nucleare (INFN) and the Dutch Nikhef, with contributions by institutions from Belgium, Germany, Greece, Hungary, Ireland, Japan, Monaco, Poland, Portugal, Spain. The construction and operation of KAGRA are funded by Ministry of Education, Culture, Sports, Science and Technology (MEXT), and Japan Society for the Promotion of Science (JSPS), National Research Foundation (NRF) and Ministry of Science and ICT (MSIT) in Korea, Academia Sinica (AS) and the Ministry of Science and Technology (MoST) in Taiwan.
\bibliography{references}
\bibliographystyle{ieeetr}
\vspace{1cm}
\textit{Verónica Villa-Ortega} \texttt{\href{mailto:veronica.villa.ortega@usc.es}{veronica.villa.ortega@usc.es}} \\
\textit{Thomas Dent} \texttt{\href{mailto:thomas.dent@usc.es}{thomas.dent@usc.es}} \\
\textit{Andrés Curiel Barroso} \texttt{\href{mailto:acurielbarroso@gmail.com}{acurielbarroso@gmail.com}}

\end{document}

%% file: table_results.tex
\begin{tabular}{l|llll|llll|llll|c}
\multicolumn{1}{c}{\textbf{Event Name}} & \multicolumn{4}{c|}{\textbf{Chirp-Mass Based}} & \multicolumn{4}{c|}{\textbf{Public Alerts}} & \multicolumn{4}{c}{\textbf{GWTC-2,-3 PE}} \\
                             & \small{BNS}  & \small{MG}    & \small{NSBH}  & \small{BBH}     & \small{BNS}  & \small{MG}   & \small{NSBH} & \small{BBH}    & \small{BNS}   & \small{MG}    & \small{NSBH} & \small{BBH} & $\mathcal{M} (M_\odot)$   \\
\hline
\rowcolor[HTML]{C0C0C0} 
GW190408\_181802              & 0     & 0     & 0     & 100     & 0     & 0     & 0     & 100     & 0     & 0     & 0     & 100   & 18.3   \\
GW190412                      & 0     & 0     & 0     & 100     & 0     & 0     & 0     & 100     & 0     & 0     & 0     & 100   & 13.3   \\
\rowcolor[HTML]{C0C0C0} 
\textbf{GW190421\_213856}     & 0     & 0     & 0     & 100     & 0     & 0     & 0     & 100     & 0     & 0     & 0     & 100   & 31.2   \\
GW190425                      & 100   & 0     & 0     & 0       & 100   & 0     & 0     & 0       & 99.5  & $<$0.1& 0     & 0     & 1.4    \\
\rowcolor[HTML]{C0C0C0} 
GW190426\_152155              & 6.0   & 40.3  & 53.7  & 0       & 57.4  & 27.6  & 15.0  & 0       & 1.1   & 29.4  & 64.2    & 0     & 2.4    \\
\rowcolor[HTML]{C0C0C0}
                              &       &       &       &         & 15.0  & 25.0  & 60.0  & 0       &       &       &       &       &        \\
GW190503\_185404              & 0     & 0     & 0     & 100     & 0     & 3.2   & 0.5   & 96.3    & 0     & 0     & 0     & 100   & 30.2   \\
\rowcolor[HTML]{C0C0C0} 
\textbf{GW190512\_180714}     & 0     & 0     & 0     & 100     & 0     & 0     & 0     & 100     & 0     & 0     & 0     & 100   & 14.6   \\
GW190513\_205428 $^{\dagger}$ & 0     & 0     & 0     & 100     & 0     & 5.2   & 0.5   & 94.3    & 0     & 0     & 0     & 100   & 21.6   \\
\rowcolor[HTML]{C0C0C0} 
GW190517\_055101              & 0     & 0     & 0     & 100     & 0     & 1.7   & $<$0.1& 98.3    & 0     & 0     & 0     & 100   & 26.6   \\
\textbf{GW190519\_153544}     & 0     & 0     & 0     & 100     & 0     & 0     & 0     & 100     & 0     & 0     & 0     & 100   & 44.5   \\
\rowcolor[HTML]{C0C0C0} 
\textbf{GW190521}             & 0     & 0     & 0     & 100     & 0     & 0     & 0     & 100     & 0     & 0     & 0     & 100   & 69.2   \\
\textbf{GW190521\_074359}     & 0     & 0     & 0     & 100     & 0     & 0     & 0     & 100     & 0     & 0     & 0     & 100   & 32.1   \\
\rowcolor[HTML]{C0C0C0} 
\textbf{GW190602\_175927}     & 0     & 0     & 0     & 100     & 0     & 0     & 0     & 100     & 0     & 0     & 0     & 100   & 49.1   \\
GW190630\_185205 $^{\dagger}$ & 0     & 0     & 0     & 100     & 0     & 5.2   & 0.5   & 94.3    & 0     & 0     & 0     & 100   & 24.9   \\
\rowcolor[HTML]{C0C0C0} 
\textbf{GW190701\_203306}     & 0     & 0     & 0     & 100     & 0     & 0     & 0     & 100     & 0     & 0     & 0     & 100   & 40.3   \\
\textbf{GW190706\_222641}     & 0     & 0     & 0     & 100     & 0     & 0     & 0     & 100     & 0     & 0     & 0     & 100   & 42.7   \\
\rowcolor[HTML]{C0C0C0} 
\textit{GW190707\_093326}     & 0     & 46.2  & 7.2   & 46.6    & 0     & 0     & 0     & 100     & 0     & $<$0.1  & 0     & $>$99.9 & 8.5    \\
\textbf{GW190720\_000836}     & 0     & 46.8  & 4.0   & 49.2    & 0     & 0     & 0     & 100     & 0     & 0.9  & 0     & 99.1 & 8.9    \\
\rowcolor[HTML]{C0C0C0} 
GW190727\_060333              & 0     & 0     & 0     & 100     & 0     & 3.0   & 0.2   & 96.8    & 0     & 0     & 0     & 100   & 28.6   \\
GW190728\_064510 $^{\dagger}$ & 0     & 47.2  & 2.4   & 50.4    & 0     & 51.6  & 14.4  &  34.0   & 0     & 1.8   & 0     & 98.2  & 8.6   \\
                              &       &       &       &         & 0     & 4.6   & 0     & 95.4    &       &       &       &       &      \\
\rowcolor[HTML]{C0C0C0} 
GW190814                      & 0     & 31.4  & 51.7  & 16.9    & 0     & 100   & 0     & 0       & 0     & 0       & 100    & 0     & 6.1     \\
\rowcolor[HTML]{C0C0C0}
                              &       &       &       &         & 0     & 0.2  & 99.8 & 0       &       &       &       &       &       \\
GW190828\_063405              & 0     & 0     & 0     & 100     & 0     & 0     & 0     & 100     & 0     & 0     & 0     & 100   & 25.0   \\
\rowcolor[HTML]{C0C0C0} 
GW190828\_065509              & 0     & 0     & 0     & 100     & 0     & 0     & 0     & 100     & 0     & 0     & 0     & 100   & 13.3   \\
GW190915\_235702 $^{\dagger}$ & 0     & 0     & 0     & 100     & 0     & 0     & 0     & 100     & 0     & 0     & 0     & 100   & 25.3  \\
\rowcolor[HTML]{C0C0C0} 
GW190924\_021846              & 0     & 30.1  & 55.7  & 14.2    & 0     & 100   & 0     & 0       & 0     & 44.8  & 3.9   & 51.3  & 5.8   \\
GW190930\_133541              & 0     & 44.0  & 13.7  & 42.3    & 0     & 100   & 0     & 0       & 0     & 8.0   & $<$0.1  & 92.0  & 8.5   \\
\hline
\hline
\rowcolor[HTML]{C0C0C0} 
\textbf{GW191105\_143521}     & 0     & 44.5  & 12.2  & 43.3      & 0     & 0     & 0     & 100     & 0     & 1.3   & $<$0.1  & 98.7  & 7.8 \\
\texttt{GW191109\_010717}     & 0     & 0     & 0     & 100       & 0     & 0     & 0     & 100     & 0     & 0     & 0     & 100   & 47.5 \\
\rowcolor[HTML]{C0C0C0} 
GW191129\_134029              & 0     & 39.3  & 28.1  & 32.6      & 0     & 0     & 0     & 100     & 0     & 4.6   & 0     & 95.4  & 7.3  \\
GW191204\_171526 $^{\dagger}$ & 0     & 47.2  & 4.2   & 48.6      & 0     & 0     & 0     & 100     & 0     & $<0.1$  & 0     & $>99.9$ & 8.6\\
\rowcolor[HTML]{C0C0C0}
GW191215\_223052              & 0    & 0     & 0      & 100       & 0     & 0     & 0     & 100     & 0     & 0     & 0     & 100   & 18.4\\
GW191216\_213338              & 0    & 47.5  & 0.6    & 51.9      & 0     & 100   & 0     & 0       & 0     & 2.0   & 0     & 98.0  & 8.3    \\
                              &      &       &        &           & 0     & 0.9  & 0     & 99.1 & & & & & \\
\rowcolor[HTML]{C0C0C0} 
GW191222\_033537 $^{\dagger}$ & 0    & 0     & 0      & 100       & 0     & 0     & 0     & 100     & 0     & 0     & 0     & 100   & 33.8  \\
GW200105\_162426              & 0    & 15.0  & 85.0   & 0         & 0     & 0     & 100   & 0       & 0     & 0.7   & 99.3   & 0     & 3.4\\
\rowcolor[HTML]{C0C0C0} 
GW200112\_155838 $^{\dagger}$ & 0    & 0     & 0      & 100       & 0     & 0     & 0     & 100     & 0     & 0      & 0     & 100   & 27.4 \\
GW200115\_042309              & 7.0  & 40.8  & 52.2   & 0         & 0     & 100   & 0     & 0       & 0.7   & 28.0   & 70.8  & 0     & 2.4  \\
\rowcolor[HTML]{C0C0C0} 
\textbf{GW200128\_022011}     & 0    & 0     & 0      & 100       & 0     & 0     & 0     & 100     & 0     & 0     & 0     & 100   & 32.0 \\
GW200129\_065458              & 0    & 0     & 0      & 100       & 0     & 0     & 0     & 100     & 0     & 0     & 0     & 100   & 27.2 \\
\rowcolor[HTML]{C0C0C0} 
\textit{GW200208\_130117}     & 0    & 0     & 0      & 100       & 0     & 0     & 0     & 100     & 0     & 0     & 0     & 100   & 27.7 \\
\textbf{GW200219\_094415}     & 0    & 0     & 0      & 100       & 0     & 0     & 0     & 100     & 0     & 0     & 0     & 100   & 27.6  \\
\rowcolor[HTML]{C0C0C0} 
\textbf{GW200224\_222234}     & 0    & 0     & 0      & 100       & 0     & 0     & 0     & 100     & 0     & 0     & 0     & 100   & 31.1 \\
\textbf{GW200225\_060421}     & 0    & 0     & 0      & 100       & 0     & 0     & 0     & 100     & 0     & 0     & 0     & 100   & 14.2  \\
\rowcolor[HTML]{C0C0C0} 
GW200302\_015811 $^{\dagger}$ & 0    & 0     & 0      & 100       & 0     & 0     & 0     & 100     & 0     & 0     & 0     & 100   & 23.4 \\
GW200311\_115853              & 0    & 0     & 0      & 100       & 0     & 0     & 0     & 100     & 0     & 0     & 0     & 100   & 26.6 \\
\rowcolor[HTML]{C0C0C0} 
GW200316\_215756 $^{\dagger}$ & 0    & 46.4  & 2.6    & 51.0      & 0     & 100   & 0     & 0       & 0     & 5.4   & $<0.1$  & 94.5  & 8.8\\
\hline
\end{tabular}

%% file: table_results_noMG.tex
\begin{tabular}{l|lll|lll|c}
\multicolumn{1}{c}{\textbf{Event Name}} & \multicolumn{3}{c|}{\textbf{Chirp-Mass Based}} & \multicolumn{4}{c}{\textbf{GWTC-2,-3 PE}} \\
                             & \small{BNS}  & \small{NSBH}  & \small{BBH} & \small{BNS} & \small{NSBH} & \small{BBH} & $\mathcal{M} (M_\odot)$   \\
\hline
\rowcolor[HTML]{C0C0C0} 
GW190425                      & 100   & 0     & 0       & 99.5 & 0     & 0      & 1.4    \\
GW190426\_152155              & 6.0   & 94.0  & 0       & 1.1   & 93.6  & 0     & 2.4    \\
\rowcolor[HTML]{C0C0C0} 
GW190707\_093326     & 0     & 7.2   & 92.8    & 0     & 0    & 100  & 8.5    \\
GW190720\_000836     & 0     & 4.0   & 96.0    & 0     & 0    & 100  & 8.9    \\
\rowcolor[HTML]{C0C0C0} \
GW190728\_064510 $^{\dagger}$ & 0     & 2.4   & 97.6    & 0     & 0    & 100   & 8.6   \\
GW190814                      & 0     & 51.7  & 48.3    & 0    & 100  & 0    & 6.1     \\
\rowcolor[HTML]{C0C0C0} 
GW190924\_021846              & 0     & 55.7  & 44.3    & 0     & 3.9    & 96.1  & 5.8   \\
GW190930\_133541              & 0     & 13.7  & 86.3    & 0    & $<$0.1  & $>$99.9 & 8.5   \\
\hline
\hline
\rowcolor[HTML]{C0C0C0} 
GW191105\_143521     & 0     & 12.2  & 87.8      & 0     & $<$0.1  & $>$99.9 & 7.8 \\
GW191129\_134029              & 0     & 28.1  & 71.9      & 0     & 0    & 100 & 7.3  \\
\rowcolor[HTML]{C0C0C0}
GW191204\_171526 $^{\dagger}$ & 0     & 4.2   & 95.8      & 0     & 0     & 100   & 8.6\\
GW191216\_213338              & 0    & 0.6    & 99.4      & 0     & 0  & 100 & 8.3    \\
\rowcolor[HTML]{C0C0C0} 
GW200105\_162426              & 0    & 85.0    & 15.0        & 0     & 99.3   & 0.7     & 3.4\\
GW200115\_042309              & 7.0  & 93.0    & 0         & 0.7     & 98.9 & 0     & 2.4  \\
\rowcolor[HTML]{C0C0C0}
GW200316\_215756 $^{\dagger}$ & 0    & 2.6   & 97.4      & 0     & $<$0.1  & $>$99.9 & 8.8\\
\hline
\end{tabular}

%% file: table_results_noz.tex
\begin{tabular}{l|llll|llll|ll}
\multicolumn{1}{c}{\textbf{Event}} & \multicolumn{4}{c|}{\textbf{Chirp-Mass Based}} & \multicolumn{4}{c|}{\textbf{Chirp-Mass Based w/o \textit{z}}} & \multicolumn{2}{c}{\textbf{\textit{z}}} \\
                                & \small{BNS}   & \small{MG}    & \small{NSBH}  & \small{BBH}       & \small{BNS}   & \small{MG}    & \small{NSBH}  & \small{BBH}   & \small{Estimated} & \small{GWTC-2,-3} \\
\hline
\rowcolor[HTML]{C0C0C0}
GW190425                       & 100   & 0     & 0     & 0       & 94.5  & 5.5   & 0     & 0       & 0.03 $\pm$ 0.01 & 0.03$^\mathrm{+0.01}_\mathrm{-0.02}$ \\
GW190426\_152155               & 6.0     & 40.3  & 53.7  & 0       & 0.3   & 37.1  & 62.5  & 0       & 0.07 $\pm$ 0.02 & 0.08$^\mathrm{+0.04}_\mathrm{-0.03}$\\
\rowcolor[HTML]{C0C0C0} 
GW190707\_093326               & 0     & 46.2  & 7.2   & 46.6    & 0     & 32.6  & 0     & 67.4    & 0.17 $\pm$ 0.04 & 0.16$^\mathrm{+0.07}_\mathrm{-0.07}$\\
GW190720\_000836               & 0     & 46.8  & 4     & 49.2    & 0     & 23.3  & 0     & 76.7    & 0.21 $\pm$ 0.05 & 0.16$^\mathrm{+0.12}_\mathrm{-0.06}$\\
\rowcolor[HTML]{C0C0C0}
GW190728\_064510 $^{\dagger}$  & 0     & 47.2  & 2.4   & 50.4    & 0     & 29.0    & 0     & 71.0      & 0.16 $\pm$ 0.04 & 0.18$^\mathrm{+0.05}_\mathrm{-0.07}$\\
GW190814                       & 0     & 31.4  & 51.7  & 16.9    & 0     & 33.4  & 45.9  & 20.7    & 0.06 $\pm$ 0.01 & 0.05$^\mathrm{+0.009}_\mathrm{-0.010}$\\
\rowcolor[HTML]{C0C0C0}
GW190924\_021846               & 0     & 30.1  & 55.7  & 14.2    & 0     & 34.0    & 44.0    & 22.0      & 0.12 $\pm$ 0.03 & 0.12$^\mathrm{+0.04}_\mathrm{-0.04}$\\
GW190930\_133541               & 0     & 44.0    & 13.7  & 42.3    & 0     & 31.1  & 0     & 68.9    & 0.23 $\pm$ 0.06 & 0.15$^\mathrm{+0.06}_\mathrm{-0.06}$\\
\hline
\hline
\rowcolor[HTML]{C0C0C0}
GW191105\_143521               & 0     & 44.5  & 12.2  & 43.3     & 0     & 35.5  & 0     & 64.5        & 0.18 $\pm$ 0.05  & 0.20$^\mathrm{+0.09}_\mathrm{-0.09}$ \\
GW191129\_134029               & 0     & 39.3  & 28.1  & 32.6     & 0     & 46.9  & 4.9   & 48.2        & 0.16 $\pm$ 0.04  & 0.16$^\mathrm{+0.05}_\mathrm{-0.06}$\\
\rowcolor[HTML]{C0C0C0} 
GW191204\_171526 $^{\dagger}$  & 0     & 47.2  & 4.2   & 48.6     & 0     & 34.1  & 0     & 65.9        & 0.14 $\pm$ 0.03  & 0.13$^\mathrm{+0.04}_\mathrm{-0.05}$\\
GW191216\_213338               & 0     & 47.5  & 0.6   & 51.9     & 0     & 38.1  & 0     & 61.9        & 0.08 $\pm$ 0.02  & 0.16$^\mathrm{+0.12}_\mathrm{-0.06}$\\
\rowcolor[HTML]{C0C0C0}
GW200105\_162426               & 0     & 15.0    & 85.0     & 0       & 0     & 17.9  & 82.1  & 0           & 0.07 $\pm$ 0.02  & 0.07$^\mathrm{+0.02}_\mathrm{-0.03}$\\
GW200115\_042309               & 7.0    & 40.8  & 52.2   & 0        & 1.1   & 37.7  & 61.2  & 0           & 0.07 $\pm$ 0.02  & 0.06$^\mathrm{+0.03}_\mathrm{-0.02}$\\
\rowcolor[HTML]{C0C0C0}
GW200316\_215756 $^{\dagger}$  & 0    & 46.4  & 2.6    & 51.0       & 0     & 17.6  & 0     & 82.4        & 0.24 $\pm$ 0.06  & 0.22$^\mathrm{+0.08}_\mathrm{-0.08}$\\
\hline
\end{tabular}